\documentclass[12pt]{revtex4}

\def\la{\mathrel{\mathpalette\fun <}}

\def\fun#1#2{\lower3.6pt\vbox{\baselineskip0pt\lineskip.9pt
\ialign{$\mathsurround=0pt#1\hfil##\hfil$\crcr#2\crcr\sim\crcr}}}

\usepackage{epsfig}

\begin{document}

\title{{\Large \bf
The impact of multipole and relativistic effects on photoionization and
radiative recombination  cross sections in hot plasmas

}}

\author{
 M. B. Trzhaskovskaya $^1$, V. K. Nikulin $^2$, and R. E. H. Clark
 $^3$\\
\em $^1$ Petersburg Nuclear Physics Institute, Gatchina  188300, Russia
 \\
$^2$ Ioffe Physical Technical  Institute,  St.Petersburg 194021,
Russia\\
 $^3$ Nuclear Data Section, International Atomic Energy Agency
Vienna A-1400, Austria}

\begin{abstract}
It is shown in the framework of the fully relativistic Dirac-Fock
treatment of photoionization and radiative recombination processes that
taking into account all significant multipoles of the radiative field is
of considerable importance at electron energy higher than several keV.
For the first time, we show that the relativistic Maxwell-Bolzmann
distribution of continuum electrons should be used in hot thermal
plasmas. This decreases the radiative recombination rate coefficient up
to several multiplies compared to the non-relativistic distribution
commonly used.
\end{abstract}

\maketitle

\hspace{9.cm} PACS number(s): 32.80.Fb; 52.20.-j

\vspace{0.5cm}

The photoionization and radiative recombination cross sections as
well as the radiative recombination rate coefficients are required
for estimates of ionization equilibria and thermal balance in
terrestrial and  astrophysical plasmas  contaminated by various ions.
In fusion reactors at temperatures above several keV, impurity atoms of
various elements may be stripped to bare nuclei. At temperatures around
1000~keV, tungsten atoms are fully stripped \cite{martin}.
In astrophysical objects such as stellar black-hole binaries and
Seyferrt galaxies, the plasma temperature may reach  150~keV~\cite{zdzi}.

At sufficiently high kinetic electron energy $E_k$ and for highly
charged ions,  multipole and  relativistic effects  should be taken
into account in calculations of photoionization cross sections (PCS),
radiative recombination cross sections (RRCS) and  radiative
recombination rate coefficients. The  effects were considered
beginning with pioneering works \cite{pratt, scof, band}.
Nevertheless, these effects are usually neglected in the application  of
these processes in plasmas  (see Refs. in \cite{AD07,badnell}). The
most extensive advanced calculations  by Badnell \cite{badnell} were
performed using the  electric dipole and semi-relativistic
approximations for electron energies  to 1.36$Z^2$~keV and for $Z \le$
54, that is up to $\sim\,$4~MeV.

We used the fully relativistic treatment of the photoionization process
in recent  calculations of PCS and RRCS for 31 ions of elements from
the range 26 $\le Z \le$ 74 \cite{AD07}.
Electron wave functions were generated in the self-consistent Dirac-Fock (DF)
framework. We took into account  all
significant multipole orders of the radiative field. Previously,
we performed   relativistic  calculations of  total and differential
RRCS for recombination of an electron with the H-, He- and Li-like
uranium ions  using this model  with regard to  the Breit
electron interaction and the main quantum electrodynamic corrections
\cite{opturan}. The influence  of the multipole effects was also
considered in our studies of the total and differential PCS
\cite{physrevb,JPb,ADNDT1}.

Exact relativistic benchmark calculations of RRCS including all
significant multipoles were carried out by Ichihara and Eichler \cite{ichihara}
for radiative recombination of the K, L and M electrons with  bare
nuclei with charge numbers $1 \le Z \le 112$ . The
results  for a  few representative cases were compared with those
derived from the widely used non-relativistic dipole  approximation in
order to assess the accuracy of the latter.

In the present paper, we discuss the influence of multipole effects on
PCS, RRCS and rate coefficients as well as the influence of  the
relativistic effects such as the relativistic transformation coefficient
between PCS and RRCS and the relativistic correction factor for the
non-relativistic rate
coefficient.  This temperature depend factor is reported  for the first
time in this paper.

The relativistic PCS  in the i-$th$ subshell per one electron can be
written in the form
\begin{eqnarray}
\sigma^{(i)}_{\rm ph} &=&
\frac{4\pi^2\alpha}{\tilde k(2j_i+1)}\sum_L\sum_\kappa\bigg[
(2L+1)Q^2_{LL}(\kappa)+LQ^2_{L+1L}(\kappa)
\\
&+& (L+1)Q^2_{L-1L}(\kappa)-2\sqrt{L(L+1)}\,Q_{L-1L}(\kappa)
Q_{L+1L}(\kappa)\bigg]\ . \nonumber
\end{eqnarray}
Here $\tilde k$ is the photon energy in $m_0c^2$, $L$ is the
multipolarity of the radiative field, $\kappa=(\ell-j)(2j+1)$,
$\ell$ and $j$ are the orbital and total angular momenta of the
electron, $\alpha$ is the fine structure constant and $Q_{\Lambda
L}(\kappa)$ is the reduced matrix element (for detailed
expressions see  \cite{AD07}). 

The cross section of the recombination process with
the capture of an electron with energy $\tilde E_k$ to the i-$th$
subshell of the ion  is expressed in terms of the
corresponding PCS as follows
\begin{equation}
\sigma^{(i)}_{\rm rr}
=Aq_i \sigma^{(i)}_{\rm ph}\, ,
\end{equation}
where $q_i$ is the number
of vacancies in the i-$th$ subshell prior to recombination. The
transformation coefficient $A$ can be derived from the principle of the
detailed balance. The exact relativistic expression for the coefficient
is written as \cite{ichihara,opturan}
\begin{equation}
A_{\rm rel} =\frac{\tilde k^2}{2\tilde E_k+\tilde E_k^2}\, ,~~~~~\tilde
E_k=\frac{E_k}{m_0c^2} .
\label{tc2}
\end{equation}

However in the majority of the RRCS  calculations, the coefficient is
used in the form
\begin{equation}
A_{\rm nrel}=\frac{k^2}{2m_0c^2 E_k} \,,~~~~~ k=m_0c^2\tilde k \, ,
\label{tc1}
\end{equation}
which may be obtained  in the non-relativistic  approximation from
Eq.~(\ref{tc2}).  The difference between $\sigma_{\rm rr}$ obtained
with Eq.~(\ref{tc2}) and Eq.~(\ref{tc1}) depends only on the electron
kinetic energy $E_k$ and can be written as

\begin{equation}
\frac{A_{\rm
nrel}-A_{\rm rel}}{A_{\rm rel}}=\frac{E_k}{2m_0c^2}
\label{diftc}
\end{equation}
\noindent
with the difference  $\sim$\,5$\%$ at $E_k$ = 50~keV  and reaching
$\sim$\,100$\%$ at $E_k$ = 1000~keV.  Consequently, at  high electron
energy, the relativistic expression (\ref{tc2}) should be used in the
RRCS calculations.

The  relativistic recombination rate coefficients  $\alpha^{(i)}_{\rm
rel}(T)$ can  be calculated using the thermal
average over RRCS. In the present paper, the
continuum  electrons are described by the relativistic
Maxwell-Boltzmann distribution function $f(E)$ normalized to unity as
follows \cite{moorad}

\begin{equation} f(E)dE=
\frac{E(E^2-1)^{1/2}}{\theta e^{1/\theta}K_2(1/\theta)} \times
e^{-(E-1)/\theta}dE.
\end{equation}

\noindent
Here $E$ is the total electron energy in units of $m_0c^2$ including the
rest energy, $\theta=k_{\beta}T/m_0c^2$ is the
characteristic dimensionless temperature, $T$ is the temperature and
$k_{\beta}$ is the Bolzmann constant. The function $K_2$ denotes the
modified Bessel function of the second order. The relativistic rate
coefficient may be written as

\begin{equation}
\alpha^{(i)}_{\rm rel}(T)=<v\sigma^{(i)}_{rr}>=
F_{\rm rel}(\theta)\cdot\alpha^{(i)}(T),
\end{equation}

\noindent
where  $v=(p/E)c$  is the electron velocity with the momentum
$p=\sqrt{E^2-1}$ and $\alpha^{(i)}(T)$ is the usual non-relativistic
rate coefficient \cite{barfield}

\begin{equation}
\alpha^{(i)}(T)=(2/\pi)^{1/2}c^{-2}(m_0k_{\beta}T)^{-3/2}q_i
\int\limits^\infty_{\varepsilon_i}k^2\sigma_{\rm ph}^{(i)}(k)
e^{(\varepsilon_i-k)/(k_{\beta}T)}dk,
\label{rc}
\end{equation}

\noindent
where  $k$ is the photon energy and  $\varepsilon_{i}$  is the binding
energy   of the i-$th$ shell.
In Eq.~(7), $F_{\rm rel}(\theta)$ is the relativistic factor
 \begin{equation}
F_{\rm rel}(\theta)=\sqrt{\frac{\pi}{2}\theta}\,\bigg/
\,K_2(1/\theta)e^{1/\theta}.
\label{fr}
\end{equation}

Using the asymptotic expansion  of the Bessel
function $K_2(1/\theta)$ at large $1/\theta$ \cite{handbook}, that is
at low temperature, we arrive at the factor $\tilde F_{\rm rel}(\theta)$,
an approximation to  $F_{\rm rel}(\theta)$

\begin{equation} \tilde F_{\rm
rel}(\theta)=1\bigg/\left(1+\frac{15}{8}\theta+
\frac{105}{128}\theta^2+...\right).
\label{frapp}
\end{equation}

\noindent
Eq.~(\ref{frapp}) provides  an  excellent
approximation for $F_{\rm rel}(\theta)$ with the terms through order
$\theta^2$ at $\theta \la 1$. The factors $F_{\rm rel}(\theta)$  and
$\tilde F_{\rm rel}(\theta)$ are compared  in Fig.~1. The solid curve
refers to the exact factor
(Eq.~(\ref{fr})) and the dashed curve refers to the approximate factor
(Eq.~(\ref{frapp})). As can be seen, there is  little  difference
between the two curves, the relative error is $\sim$\,4\% at
$k_{\beta}T$ = 500~keV and 25\% at $k_{\beta}T$ = 1000~keV.

\vspace{-1.0cm}


\begin{figure}[h]
\centerline{ \epsfxsize=7 cm\epsfbox{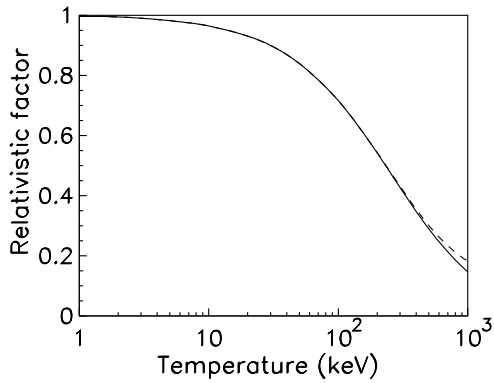}}

\vspace{-1.cm}

\caption{\small  Exact  factor $F_{\rm rel}(T)$ (solid)
and approximate  factor $\tilde F_{\rm rel}(T)$ (dashed).
}
\end{figure}

As is seen from Fig.~1, the use of the relativistic distribution
instead of  non-relativistic one results in a decrease of rate
coefficients by a factor of 1.2 at plasma temperature
$k_{\beta}T$ =~50~keV and up to factor of 7 at $k_{\beta}T$ = 1~MeV.

Let us next consider the influence of the multipole effects.
 The electric dipole approximation takes into account only terms
with $L$=1  in Eq.~(1).  As is well known, the dipole  approximation
holds at a low electron energy $E_k$ but  breaks down at a higher
energy.

\vspace{-1.cm}
\begin{figure}[h]
\centerline{ \epsfxsize=9cm\epsfbox{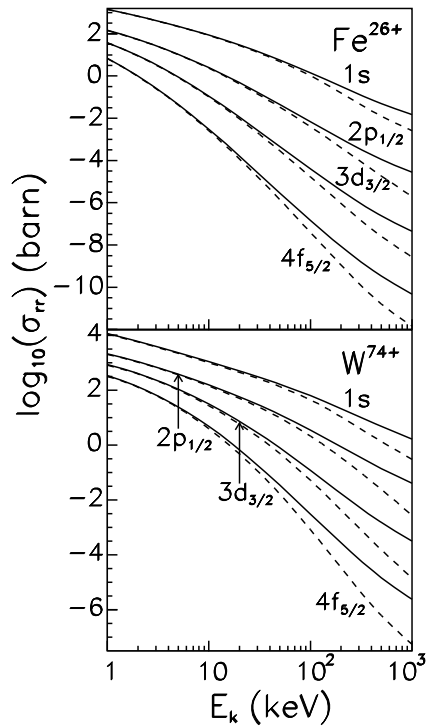}}

\vspace{-1.cm}

\caption{\small Subshell RRCS (in barns)  calculated taking into
account all multipoles $L$ (solid) and in the dipole
approximation (dashed). }

\end{figure}

In Fig.~2, we compare RRCS
obtained in the dipole approximation $\sigma_{\rm {rr}}$(dip)
(dashed curves) with RRCS calculated with  all
multipoles $\sigma_{\rm {rr}}$(L) making significant contribution
(solid curves) for bare
nuclei of two representative elements Fe ($Z$=26) and W ($Z$=74).  The
energy range under consideration is 1~keV $\le E_k \le$ 1000~keV.
As is seen in Fig.~2, the curves begin to diverge
noticeably even at several keV.  At the highest energy 1000~keV,  in
the case of W$^{74+}$, $\sigma_{\rm {rr}}$(dip) is smaller than the
exact value $\sigma_{\rm {rr}}$(L) by a factor of $\sim$\,5 for the
$1s$ shell and by a factor of $\sim$\,40 for the $4f_{5/2}$
subshell. Our calculations showed that  the  relative difference
between the exact calculation of RRCS and the dipole approximation
 \begin{equation}
\Delta_{\rm
RRCS}=\frac{\sigma_{\rm rr}{\rm(L)}-\sigma_{\rm rr}\rm(dip)}
{\sigma_{\rm rr}\rm(L)}\cdot 100\%
\label{difdip}
\end{equation}

\noindent
varies in the range  $\sim$\,3-20\%  for  shells with  different
orbital momenta $\ell_i$ at $E_k$=10~keV, $\sim$\,15-50\% at
$E_k$=50~keV and reaches  several multiples at $E_k$=1000~keV. The
dependence of $\Delta_{\rm RRCS}$ on  $\ell_i$ is shown to be
considerable, $\Delta_{\rm RRCS}$ being larger with increasing
$\ell_i$. The difference $\Delta_{\rm RRCS}$ was found to increase with
$Z$, especially for  high energy.

Calculations show that to achieve  accuracy $\sim$\,0.01\% in the
PCS, {\em e.g.}, for the $1s$ and $3d$ shells of the ion W$^{73+}$ one
has to take into account  all terms in Eq.~(1) up  to $L$=5  and $L$=8,
respectively at $E_k$=10~keV, to $L$=7 and 11 at $E_k$=100~keV and
to $L$=19 and 31  at $E_k$=1000~keV.

\noindent {\bf Table I.} Comparison of our  PCS
 with results by  Badnell \cite{badnell}
for the $1s$ shell of the  H-like ion   Xe$^{53+}$.
$\Delta_{\rm PCS}=\bigg[ [\sigma_{\rm ph}(\rm present)-\sigma_{\rm
ph}(\rm Badnell)]/\sigma_{\rm ph}(\rm present)\bigg]\cdot 100\%$.

\tabcolsep=0.2cm

\begin{center}
\begin{tabular}{cccc} \hline \hline

&\multicolumn{2} {c}  {$\sigma_{\rm ph}$, Mb} & \\
\cline{2-3}

$E_k$, keV  &  Badnell &  Present & $\Delta_{\rm PCS},\%$  \\
\hline
 0.00083 & 2.246(-3) & 1.937(-3)&-16 \\
 0.03967 & 2.240(-3) & 1.935(-3)&-16 \\
 0.3967 & 2.186(-3) & 1.892(-3)&-16 \\
   3.967   & 1.734(-3) & 1.523(-3)&-14 \\
   39.67   & 3.256(-4) & 3.114(-4)&-4.5 \\
   83.31   & 9.095(-5) & 9.206(-5)&1.2  \\
 182.4   & 1.539(-5) & 1.740(-5)&12   \\
  396.7   & 1.894(-6) & 2.802(-6)&32  \\
  833.1   & 2.071(-7) & 5.495(-7)&62 \\
 1824.   & 1.730(-8) & 1.318(-7)&87  \\
3967.   & 1.350(-9) & 4.117(-8)&97    \\
\hline \hline

\end{tabular}
\end{center}

In Table I, we compare our present PCS calculations with the
corresponding results of Badnell~\cite{badnell} for the $1s$ shell
of the H-like ion Xe$^{53+}$. The case of the  one-electron ion is
particularly convenient for checking  the influence of the higher
multipoles and the method of calculation because there are no any
inter-electron interactions.  In this case,  the PCS must be
independent of the gauge used in calculations for correct wave
functions.

We found that our calculation is in excellent agreement with
values from \cite{ichihara} where all multipoles $L$ were involved.
For the $1s$ shell of the H-like ion Xe$^{53+}$, the two calculations
coincide  with an accuracy of the three significant digits presented in
\cite{ichihara} in the wide energy range  1~eV $\le E_k \le$ 6000~keV. By
contrast, PCS obtained by Badnell exceed our values and results from
\cite{ichihara} by $\sim$\,16\% in the energy range $E_k \la$~4~keV
and diminish progressively at higher energies becoming lower by a
factor of $\sim$\,8 at $E_k \approx$~1800~keV and a factor of
$\sim$\,30 at $E_k \approx$~4000~keV  compared with our values. The comparison
of our PSC values and results from \cite{ichihara} with calculation by
Badnell \cite{badnell} for the lighter ion  Fe$^{23+}$ reveals a
similar tendency, but smaller in magnitude.

The reason of the difference at  low energies is
unclear for us because the non-dipole terms make a small
contribution at low energies (see Fig.~2). It is possible that the
difference arises from the  methods of calculation used in
\cite{badnell}. The difference at high energies ($>100$~keV)
must be due to neglect of the higher multipoles  and possibly also
due to the semi-relativistic approximation adopted in~\cite{badnell}.

\begin{figure}[h]
\centerline{ \epsfxsize=7cm\epsfbox{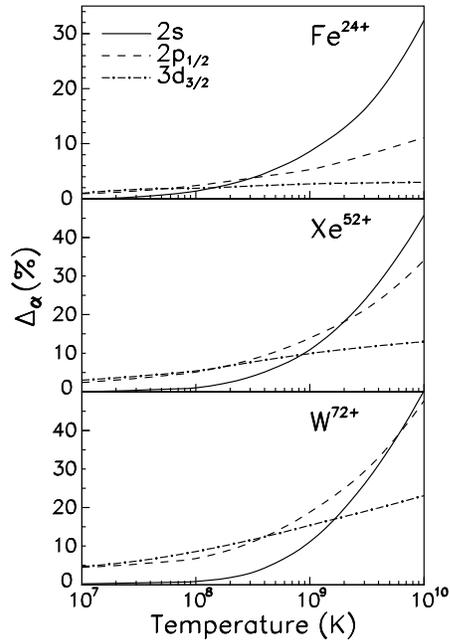}}

\vspace{-0.5cm}

\caption{\small Difference $\Delta_{\alpha}$  between rate coefficients calculated
 with using all
multipoles and in the dipole approximation for the $2s$ (solid),
$2p_{1/2}$ (dashed),  and $3d_{3/2}$ (dash-dotted) shells. }
\end{figure}

From the  discussion above, it would  be expected that the dipole
approximation would also fail in calculations of
rate coefficients  at a high temperature $T$. In Fig.~3, we
present the difference $\Delta_{\rm \alpha}$ between the exact
$\alpha^{(i)}(\rm L)$ and the dipole $\alpha^{(i)}(\rm dip)$ values of
partial rate coefficients. The difference is defined in the same way as in
Eq.~(\ref{difdip}). The difference $\Delta_{\rm \alpha}$ is given for
the $2s, 2p_{1/2}$ and $3d_{3/2}$ electrons recombining with the
He-like ions Fe$^{24+}$, Xe$^{52+}$ and W$^{72+}$. These shells are the
lowest ones making a  large contribution to the total rate
coefficients. As is evident from the figure, the difference
$\Delta_{\rm \alpha}$ is larger for the heavy, highly charged  ions.
The inclusion of higher multipoles may change partial rate coefficients
by $\sim$\,7\% at temperature $T=10^8$~K, by $\sim$\,20\% at $T=10^9$~K
and by $\sim$\,50\% at $T=10^{10}$~K for W$^{72+}$. This means that
total rate coefficients obtained within the dipole approximation have
to be considerably smaller than the accurate values obtained with
regard to all multipoles $L$.

In conclusion  we have clearly demonstrated the
importance of multipole effects to the PCS, RRCS and rate coefficient calculations
at electron energy of the order of 10~keV and higher. We have also
showed that in hot plasmas, the relativistic Maxwell-Bolzmann
distribution of continuum  electrons must be used in the rate
coefficient calculations. It should be noted that the relativistic
rates  may be similarly  obtained for processes of dielectronic
recombination and the electron impact ionization in hot plasmas.

This work was funded through International Atomic Energy Agency under
Contract No.\,13349/RBF and partially by Russian Foundation for
Basic Research (project No.\,06-02-16489) which are gratefully
acknowledged.

\end{document}